# Reducing decision errors in the paired comparison of the diagnostic accuracy of screening tests with Gaussian outcomes

Brandy M Ringham[1*], Todd A Alonzo[2], John T Brinton[1], Sarah M Kreidler[1], Aarti Munjal[1], Keith E Muller[3], and Deborah H Glueck[1]

[1]Department of Biostatistics and Informatics, Colorado School of Public Health, University of Colorado, Anschutz Medical Campus, Aurora, Colorado 80045, U.S.A.

[2]Department of Preventive Medicine, University of Southern California, Los Angeles, California 90089, U.S.A.

[3]Department of Health Outcomes and Policy, University of Florida, Gainesville, FL 32611, U.S.A.

[*]Correspondence: brandy.ringham@ucdenver.edu

Email:

| | |
|---|---|
| Brandy M Ringham | brandy.ringham@ucdenver.edu |
| Todd A Alonzo | talonzo@childrensoncologygroup.org |
| John T Brinton | john.brinton@ucdenver.edu |
| Sarah M Kreidler | sarah.kreidler@ucdenver.edu |
| Aarti Munjal | aarti.munjal@ucdenver.edu |
| Keith E Muller | kmuller@ufl.edu |
| Deborah H Glueck | deborah.glueck@ucdenver.edu |

## Abstract

**Background:** Scientists often use a paired comparison of the areas under the receiver operating characteristic curves to decide which continuous cancer screening test has the best diagnostic accuracy. In the paired design, all participants are screened with both tests. Participants with unremarkable screening results enter a follow-up period. Participants with suspicious screening results and those who show evidence of disease during follow-up receive the gold standard test. The remaining participants are classified as non-cases, even though some may have occult disease. The standard analysis includes all study participants in the analysis, which can create bias in the estimates of diagnostic accuracy. If the bias affects the area under the curve for one screening test more than the other screening test, scientists may make the wrong decision as to which screening test has better diagnostic accuracy. We propose a weighted maximum likelihood bias correction method to reduce decision errors.

**Methods:** Using simulation studies, we assessed the ability of the bias correction method to reduce decision errors. The simulations compared the Type I error rate and power of the standard analysis with that of the bias-corrected analysis. We varied four factors in the simulation: the disease prevalence, the correlation between screening test scores, the rate of interval cases, and the proportion of cases who receive the gold standard test. We demonstrate the proposed method with an application to a hypothetical oral cancer screening study.

**Results:** The amount of bias in the study and the performance of the bias correction method depend on characteristics of the screening tests and the disease, and on the percentage of study participants who receive the gold standard test. In studies with a large amount of bias in the difference in the full area under the curve, the bias correction method reduces the Type I error rate and improves power for the correct decision.

**Conclusion:** The bias correction method reduces decision errors for some paired screening trials. In order to determine if bias correction is needed for a specific screening trial, we recommend the investigator conduct a simulation study using our software.

**Keywords:** cancer screening, differential verification bias, area under the curve, type I error, power, paired screening trial, receiver operating characteristic analysis

## Background

Paired screening trials are common in cancer screening. For instance, one of the designs considered for a planned oral cancer screening study was a paired comparison of the oral tactile visual exam with the VELscope imaging device [1]. Two recent breast cancer screening studies used a paired design to compare film and digital mammography [2, 3]. In paired cancer screening trials, investigators screen all participants with both screening tests. Participants with unremarkable screening test scores on both tests enter a follow-up period. Participants with suspicious screening test scores or who show signs and symptoms of disease during the follow-up period, undergo further workup leading to the gold standard test, biopsy. The remaining participants are assumed to be disease-free. In truth, they may have occult disease.

In the trial by Lewin *et al.* [2], the investigators used the standard analysis to compare the full areas under the receiver operating characteristic curves. The standard analysis includes all participants, even those whose disease status is not verified with the gold standard test. Because some cases are misclassified as non-cases, the receiver operating characteristic curves and the corresponding areas under the curves may be biased, thus causing decision errors [4]. We use bias in the epidemiological sense of the word as the difference between what the study investigator observes and the true, unobservable state of nature [5]. If the bias is severe enough, investigators can detect a difference between screening tests when there is none, or decide that the tests are different, but conclude incorrectly that the inferior test is superior. Choosing the inferior screening test can delay diagnosis of cancer and increase morbidity and mortality.

We propose a bias-corrected hypothesis test to reduce decision errors in paired cancer screening trials. Under the assumption that the screening test scores follow a bivariate Gaussian distribution, conditional on disease status, we use an iterative, maximum likelihood approach to reduce the bias in the estimates of the mean, variance, and correlation. The resulting estimates are then used to reduce bias in the estimates of the diagnostic accuracy of the screening tests.

Screening trials are subject to different biases depending on the choice of reference standard and analysis plan [6 − 9]. Imperfect reference standard bias occurs when the investigator evaluates all study participants with an imperfect reference standard [7, 8, 10]. Partial verification bias occurs if 1) the investigator evaluates only some participants with a gold standard, 2) the decision to use the gold standard test depends on the screening test outcome, and 3) the investigator only includes participants with gold standard results in the analysis, a complete case analysis [11]. Differential verification bias occurs when 1) the investigator uses a gold standard test for a portion of participants and an imperfect reference standard for the remaining participants, 2) the decision to use the gold standard test depends on the results of the screening test, and 3) the investigator includes data from all participants in the analysis [8]. Paired screening trial bias [4], the focus of our research, is a special case of differential verification bias. Paired screening trial bias occurs in paired screening trials when three conditions hold: 1) The investigator analyzes data from all the participants 2) the screening tests are subject to differential verification bias and 3) each screening test has associated with it a different threshold score that leads to the gold standard test [4].

In the following sections we describe the bias correction method and evaluate its performance by simulation. In the Methods section, we explain the study design of interest, outline the assumptions and notation, delineate the bias correction algorithm, and describe the design of the simulation studies. In the Results section, we report on the results of the simulation studies and demonstrate the utility of the method using a hypothetical oral cancer screening study. Finally, in the Discussion section, we discuss the implications of our method.

## Methods

### Study Design

The study design of interest is a paired study of two continuous cancer screening tests. The design was considered by Lingen [1] in a planned oral cancer screening trial, and was used in two recent breast cancer screening trials [2, 3]. A flowchart of the study design is shown in Figure 1.

We consider the screening studies from two points of view [12]. We consider the point of view of the omniscient observer who knows the *true* disease status for each participant. We also consider the point of view of the study investigator, who can only see the *observed* results of the study.

The study investigator determines disease status in two ways: through follow-up, or through a definitive method for the ascertainment of disease status. Any score that exceeds the threshold of suspicion defined for each screening test triggers the use of a gold standard test. Cases identified due to remarkable screening test scores are referred to as *screen-detected* cases. Participants with unremarkable screening test scores on both screening tests enter a follow-up period. Some participants may show signs and symptoms of disease during the follow-up period, leading to a gold standard test. Participants who are diagnosed as cases because of signs and symptoms during the follow-up period are referred to as *interval* cases. We refer to the collection of screen-detected cases and interval cases as the *observed* cases. Participants with unremarkable screening test scores who do not show signs and symptoms of disease during the follow-up period are assumed to be disease-free.

Under the assumption that the gold standard test is 100% sensitive and specific, the study design described above will correctly identify all non-cases. However, the design may cause some cases to be missed. *Missed* cases occur when study participants who actually have disease receive unremarkable screening test scores and show no signs or symptoms of disease.

We present a graph of a hypothetical dataset of screening test scores (Figure 2) to illustrate how the study investigator *observes* disease status. The axes represent the thresholds of suspicion for each screening test. The study investigator *observes* every case in the gray region, but a smaller portion of cases in the white region. We can identify the missed cases because we present this graph from an omniscient point of view.

**Standard Analysis**

In the standard analysis, the study investigator compares the diagnostic accuracy of the two screening tests, measured by the full area under the receiver operating characteristic curve. The receiver operating characteristic curves are calculated using data from all cases and non-cases *observed* in the study. When cases are missed, the study investigator calculates the sensitivity and specificity of the screening test incorrectly. The error in sensitivity and specificity causes concomitant errors in the area under the curve. Thus, the *observed* area under the curve can differ from the *true* area under the curve.

The error in sensitivity may be large [4]. However, the error in specificity is negligible. Thus, our proposed bias correction method only corrects the estimation of the sensitivity and does not correct specificity. In fact, a specificity correction is typically not possible, because very few non-cases actually receive the gold standard test.

**Assumptions, Definitions, and Notation**

We make a series of assumptions. Let $n$ be the total number of study participants, and $\pi$ the prevalence of disease in the population. Assuming simple random sampling, the number of participants with disease is $M \in [0, n]$, and is distributed

$$M \sim \text{Binomial}(n, \pi). \tag{1}$$

Let $i \in \{1, 2, \ldots, n\}$ index participants, $j \in \{1, 2\}$ index the screening test, and $k \in \{0, 1\}$ indicate the presence ($k = 1$) or absence ($k = 0$) of disease. Let $X_{ijk}$ denote the screening test score for the $i^{\text{th}}$ participant on the $j^{\text{th}}$ screening test with *true* disease status $k$. For each disease status, $k$, and participant $i$, the pair of test scores, $X_{i1k}$ and $X_{i2k}$, are independently and identically distributed bivariate Gaussian random variables with means, $\mu_{jk}$, variances, $\sigma^2_{jk}$, and correlation, $\rho_k$.

Let $a_j$ be the threshold of suspicion for screening test $j$. All scores above the threshold will trigger the use of a gold standard test. Let $I$ be the event that a participant shows signs and symptoms of disease during a follow-up period and $P(I|k = 1) = \psi$. We assume that $P(I|k = 0) = 0$.

For screening test $j$, the *percent ascertainment* is 100 times the number of participants with disease who score above the threshold on screening test $j$, divided by the total number of participants with *observed* disease.

**Bias Correction Algorithm**

We describe an algorithm to reduce bias in estimates of the area under the curve. The algorithm requires four steps.

**Step 1. Partition.** The *observed* cases can be stratified into two sets, shown in Figure 2. Let $A$ (data in the gray area) be the set of *true* cases with at least one screening test score above its respective threshold. Let $B$ (data in the white area) be the set of *true* cases where the scores on both screening tests fall below their respective thresholds. The percentages of participants with *observed* disease in sets $A$ and $B$ differ: all cases of disease in set $A$ are *observed*, but only a fraction of cases are *observed* in set $B$. We handle the estimation for each set separately, and combine the estimates using weighting proportional to the sampling fraction [13; Equation 3.3.1, p. 81].

**Step 2. Maximum Likelihood Estimation.** We obtain maximum likelihood estimates of the bivariate Gaussian parameters for the cases. The estimation process follows an iterative method suggested by Nath [14]. The method allows unbiased estimation of bivariate Gaussian parameters from sample spaces truncated to be convex. Set $A$ is not a convex set. To obtain convex sets, we further partition the sample space into quadrants $Q_l$, $l \in \{1, 2, 3, 4\}$, as shown in Table 1.

The starting values for the iteration are the sample statistics for the *observed* cases in each quadrant. From the four quadrant specific estimates, we choose the set that maximizes the log likelihood of the full bivariate Gaussian distribution with respect to the *observed* data. We refer to that set as the Nath estimates, denoted by $\hat{\mu}_{11,N}$, $\hat{\mu}_{21,N}$, $\hat{\sigma}^2_{11,N}$, $\hat{\sigma}^2_{21,N}$, and $\hat{\rho}_{1,N}$. Note that the $k$ index is always equal to one since we only estimate parameters for the bivariate Gaussian distribution of cases.

We require the sample variance as a starting value for the Nath algorithm. However, we can only calculate the sample variance for quadrants with two or more observations. Thus, if $Q_l$ has only one observation, we do not calculate the quadrant specific estimates for that quadrant.

**Step 3. Weighting.** The Nath estimates are only based on one quadrant of data. We use the weighting described below to incorporate data from all quadrants and thereby lower the variance. We refer to the weighted estimates as the *corrected* estimates and use them to calculate the *corrected* areas under the receiver operating characteristic curves. If either set $A$ or set $B$ contain only one observation, we do not conduct the weighting and instead use the Nath estimates as the *corrected estimates*.

The Nath estimates are used as inputs for calculating the sampling fraction for sets $A$ and $B$. Define the estimated probability of $A$ as

$$\widehat{\lambda} = 1 - \Phi\left(\frac{a_1 - \widehat{\mu}_{11,N}}{\widehat{\sigma}_{11,N}}, \frac{a_2 - \widehat{\mu}_{21,N}}{\widehat{\sigma}_{21,N}}, \widehat{\rho}_{1,N}\right). \tag{2}$$

Let $k' = 1$ if a participant is *observed* to have disease and 0 otherwise. For the screen positive cases, let $\overline{X}_{jk',A}$ be the sample mean, $S_{jk',A}$ be the sample standard deviation for screening test $j$, and $r_{k',A}$ be the sample correlation between the screening tests. For the interval cases, let $\overline{X}_{jk',B}$ be the sample mean, $S_{jk',B}$ be the sample standard deviation for screening test $j$, and $r_{k',B}$ be the sample correlation between the screening tests. Let $\widehat{\mu}_{j1,W}, \widehat{\sigma}^2_{j1,W}$, and $\widehat{\rho}_{1,W}$ be the weighted estimates of the mean, variance, and correlation for the screening test scores for the cases.

We derived expressions for the weighted parameter estimates using the conditional covariance formula [15; Proposition 5.2, p. 348] and the definition of the weighted mean [13; Equation 3.2.1, p. 77]. We define the estimates as follows:

$$\widehat{\mu}_{11,W} = \widehat{\lambda}\overline{X}_{11,A} + \left(1 - \widehat{\lambda}\right)\overline{X}_{11,B}, \tag{3}$$

$$\widehat{\mu}_{21,W} = \widehat{\lambda}\overline{X}_{21,A} + \left(1 - \widehat{\lambda}\right)\overline{X}_{21,B}, \tag{4}$$

$$\widehat{\sigma}^2_{11,W} = \mathcal{G}_1 + \mathcal{H}_1 - \widehat{\mu}^2_{11,W}, \tag{5}$$

$$\widehat{\sigma}^2_{21,W} = \mathcal{G}_2 + \mathcal{H}_2 - \widehat{\mu}^2_{21,W}, \tag{6}$$

and

$$\widehat{\rho}_{1,W} = \widehat{\sigma}^2_{11,W}\widehat{\sigma}^2_{21,W}\left(\mathcal{P} + \mathcal{Q} - \widehat{\mu}_{11,W}\widehat{\mu}_{21,W}\right) \tag{7}$$

where

$$\begin{aligned}
\mathcal{G}_j &= \widehat{\lambda}\left(\overline{X}^2_{j1,A} + S^2_{j1,A}\right), \\
\mathcal{H}_j &= \left(1 - \widehat{\lambda}\right)\left(\overline{X}^2_{j1,B} + S^2_{j1,B}\right), \\
\mathcal{P} &= \widehat{\lambda}\overline{X}_{11,A}\overline{X}_{21,A} + \widehat{\lambda}S_{11,A}S_{21,A}r_{1,A},
\end{aligned}$$

and

$$\mathcal{Q} = \left(1 - \widehat{\lambda}\right)\overline{X}_{11,B}\overline{X}_{21,B} + \left(1 - \widehat{\lambda}\right)S_{11,B}S_{21,B}r_{1,B}.$$

**Evaluation of Bias Correction**

We compared the Type I error and power of the *observed*, *corrected*, and *true* analyses. For the *observed* analysis, we used the *observed* disease status as inputs for the paired comparison of the diagnostic accuracy of the screening tests. The *observed* analysis replicates the standard analysis performed by the study investigator of a cancer screening trial. For the *corrected* analysis, we used the proposed bias correction approach. Finally, both the *observed* and *corrected* analyses were compared to the *true* analysis. The *true* analysis assumes that the study investigator knows the *true* disease status of every participant.

For each analysis, we tested the null hypothesis that there is no difference in the areas under the receiver operating characteristic curves. We calculated the areas under the binormal receiver operating characteristic curves [16; p. 122]. We then calculated the variance of the difference in the areas under the curves and conducted a hypothesis test using the method of Obuchowski and McClish [17]. Ethically, we can only conduct a screening trial if we have clinical equipoise, *i.e.*, if we do not have evidence to favor one screening strategy over the other. Thus, we used a two-sided hypothesis test.

To assess screening test performance, we compared the Type I error and power for the *observed*, *corrected,* and *true* analyses. Power is the probability that we reject the null hypothesis. When the hypothesis test detects a difference between the screening tests, the study investigator uses the estimates of diagnostic accuracy to decide which screening test to implement. However, the estimates of diagnostic accuracy can be biased. Consequently, even though the study investigator correctly concludes that there is a difference between the two screening tests, the investigator may incorrectly choose to implement the screening test with the lower diagnostic accuracy. To quantify the decision error, we divided power into the *correct rejection fraction* and the *wrong rejection fraction*. The *correct rejection fraction* is the probability that the hypothesis test rejects and the screening test with the larger *observed* area under the curve is the screening test with larger *true* area under the curve. The *wrong rejection fraction* is the probability that the hypothesis test rejects but the screening test with the larger *observed* area under the curve is the screening test with the smaller *true* area under the curve.

**Design of Simulation Studies under the Gaussian Assumption**

We conducted simulations under the assumptions listed in Assumptions, Definitions, and Notation. We considered two states of nature; one where the null hypothesis holds and one where the alternative hypothesis holds. When the null holds, the two screening tests have the same diagnostic accuracy. When the alternative holds, the screening tests have different diagnostic accuracies.

The simulation studies varied four factors: 1) the disease prevalence, 2) the proportion of cases that exhibit signs and symptoms of disease during follow-up, 3) the correlation between Test 1 and 2 scores, and 4) the thresholds that trigger a gold standard test. The four factors change the amount of bias in the estimates of the diagnostic accuracy of each screening test. In addition, the factors change the number and proportion of *observed* cases. For brevity, we do not present the correlation study results in this manuscript. The correlation between the screening tests had a negligible effect on the performance of the bias correction method.

For each combination of parameter values, we simulated paired screening test scores and a binary indicator of *true* disease status. Based on the described study design, we deduced the *observed* disease status. We calculated the *true*, *observed*, and *corrected* areas under the curves and assessed the results of the simulation study using the metrics introduced in the Evaluation of Bias Correction section. We used 10,000 realizations of the simulated data to ensure that the error in the estimation of probabilities occurred in the second decimal place.

**Design of Non-Gaussian Simulation Studies**

We conducted a second set of simulation studies to examine the robustness of the bias correction method to deviations from the Gaussian assumption. We considered two possible deviations: zero-weighted data, which corresponds to screening test scores with zero values, and a multinomial distribution, which corresponds to reader preferences for a subset of the possible screening test scores. We evaluated the performance of the bias correction method as described in the Evaluation of Bias Correction section.

To generate the zero-weighted data where the occurrence of zeroes is correlated between the two screening tests, we generated two sets of correlated Bernoulli random variables, one for cases and one for non-cases, so that the probability that the score on Test 1 is zero is $p_{1k}$, the probability that the score on Test 2 is zero is $p_{2k}$ and the probability that both screening test scores are zero is $q_k$. If the Bernoulli random variable was one, we replaced the associated screening test score with a zero. Otherwise, the screening test score remained as it was. We set $p_{jk}$ equal to a range of values between 0.01 and 0.90, creating a series of datasets. The marginal probabilities put constraints on the possible values for $q_k$ [18]. We set $q_k$ to the median allowed agreement for each pairing of $p_{jk}$.

To generate the multinomial data, we binned the bivariate Gaussian data. We created a series of datasets, each with a different bin size. The bin sizes were $1/10$, $1/2$, 1, 2, and 5 times the variance. We set the disease prevalence to 0.01 (low), 0.14 (medium) and 0.24 (high).

**Results**

**Effect of Disease Prevalence**

Changing the disease prevalence affects both the case and non-case case sample size. In turn, this changes the number of observations used to calculate sensitivity and specificity, which affects the Type I error rate and power of the analysis.

**Choice of Parameter Values.** The diagnostic accuracy of the screening tests are similar to those found in the study by Pisano *et al.* [3]. When the null holds, we fixed the *true* area under both curves to be 0.78. When the alternative holds, we fixed the *true* area under the curve to be 0.78 for Test 1 and 0.74 for Test 2 for a *true* difference of 0.04. The sample size was fixed at 50,000 and the rate of signs and symptoms at 0.10. The thresholds of suspicion were set to confer a large degree of paired screening trial bias.

We varied the disease prevalence between 0.01 and 0.24. The choice reflects cancer rates seen in published cancer studies (prevalence of prostate cancer in Tobago men aged 70-79 is 0.28 in [19]; rate of breast cancer is roughly 6 per 1000 in [2] and [3]; prevalence of oral lesions is 0.14 in [20]; prevalences of all major cancers conditional on age and gender fall between 0 and 0.17 per SEER Tables 4.25, 5.15, 15.27, 16.21, 20.27, 23.15 [21]).

**Type I Error.** As shown in panel A of Figure 3, in the presence of paired screening trial bias, the *observed* Type I error rate ranges between 0.06 and 0.99. The *corrected* Type I error rate ranges between 0.03 and 0.08. Higher disease prevalence has higher Type I error rates.

**Power.** In panels B and C of Figure 3, the *corrected* analysis has a high correct rejection fraction across all disease prevalences likely to be encountered in a cancer screening study. The correct rejection fraction ranges between 0.58 and 0.96. By contrast, the *observed* analysis often wrongly concludes that the worst screening test is best. For the example shown, the correct rejection fraction for the *observed* analysis is zero while the wrong rejection fraction ranges between 0.86 and 1.00.

**Effect of the Rate of Signs and Symptoms**

When the rate of signs and symptoms is low, there is a higher potential for occult disease and, hence, error in the estimates of diagnostic accuracy.

**Choice of Parameter Values.** We defined the fixed parameter values as in the Effect of Disease Prevalence section. We varied the rate of signs and symptoms across a range of clinically relevant values. In cancer screening, the rate of signs and symptoms is not typically reported, but can be approximated using surrogates. In the trial by Lewin *et al.* [2], we can use the proportion of interval cancers out of all *observed* cancers as an estimate of the largest possible rate of signs and symptoms. An earlier breast cancer screening trial by Bobo, Lee and Thames [22] evaluated the performance of clinical breast exams for early identification of breast cancer. If we assume an abnormal clinical breast exam is the same as showing signs and symptoms of disease, then having an abnormal breast exam but a benign mammogram is akin to a woman screening negative on mammography but then showing signs and symptoms of disease. Based on these surrogates, the approximate values for the rate of signs and symptoms were 0.11 [2] and 0.05 [22]. We chose a range of 0 to 0.20 for our simulation studies.

**Type I Error.** In panel A of Figure 4, Type I error declines with increasing rate of signs and symptoms. The Type I error rate of the *corrected* analysis is below nominal at low disease prevalence, and ranges from 0.03 to 0.14 at medium and high disease prevalence. Type I error rate of the *observed* analysis ranges between 0.05 and 0.07 at low disease prevalence, and 0.77 to 1.00 at medium and high disease prevalence.

**Power.** In panels B and C of Figure 4, increasing the rate of signs and symptoms improves the correct rejection fraction. The correct rejection fraction for the *true* analysis is 0.77 at low disease prevalence and 1.00 at medium and high disease prevalence. The correct rejection fraction of the *corrected* analysis ranges from 0.56 to 0.59 at low disease prevalence and 0.72 to 0.97 at medium and high disease prevalence. The *observed* analysis has a correct rejection fraction of zero, while the wrong rejection fraction ranges from 0.51 to 1.00. Under the extreme bias conditions of the simulation, the study investigator using the *observed* results will either incorrectly decide that the worst screening test is best, or conclude there is no difference between the two screening tests.

**Effect of Percent Ascertainment**

The threshold of suspicion for each screening test determines what percentage of cases receive the gold standard test. We chose a threshold for each screening test that corresponds to around 15%, 50% and 80% ascertainment. Percentages are approximate because the case numbers are discrete.

Paired screening trial bias occurs when the percent ascertainment is different for the two screening tests. When the percent ascertainment is equivalent for the two screening tests, estimates of diagnostic accuracy for each screening test may still be subject to differential verification bias. However, since each test is biased by the same amount, the difference in the areas under the curves will remain unchanged.

Table 2 shows the Type I error of the *true*, *observed* and *corrected* analyses for a selection of percent ascertainment levels. We do not discuss the power results since the Type I error of the *observed* analysis is so high and power is bounded below by Type I error rate.

**Choice of Parameter Values.** We defined the fixed parameter values as in the Effect of Disease Prevalence. We chose values for the thresholds that allowed the amount and source of the bias (Test 1 or Test 2) to vary across a wide range of possible screening study scenarios.

**Type I Error.** In general, when the study is biased, the Type I error rate of the *observed* analysis is too high (0.23 to 1.00). The Type I error rate of the *corrected* analysis is closer to nominal than that of the *observed* analysis, but is also too high (0.12 to 0.95). For the pairings with no paired screening trial bias, the *observed* analysis has lower than nominal Type I error rates (0.02) while the *corrected* analysis has Type I error rates up to 0.26. When the percent ascertainment is high (80/80) for both screening tests, the Type I error rate of the *corrected* analysis is below nominal.

**Robustness to Non-Gaussian Data**

Although the bias correction method was developed under an assumption that the data were bivariate Gaussian, screening data may not follow the Gaussian distribution. Multinomial and zero-weighted data are common in imaging studies. Often, in imaging studies, readers give the image a score of zero to indicate that no disease is seen, resulting in a data set where multiple values are zero. In addition, reader preferences for a subset of scores can produce a multinomial data set.

The bias correction method can still be used if data are multinomial or zero-weighted. For multinomial data, the method performs well when the bins for the multinomial are scaled to less than half the variance. For zero-weighted data, the bias correction algorithm can reduce decision errors when up to 30% of the non-cases and 1% of the cases receive a score of zero. The study investigator can obtain a rough estimate of the percentage of cases that receive zero scores from the published rate of interval cancers.

**Demonstration**

Figure 5 shows the receiver operating characteristic curves for a hypothetical oral cancer screening trial similar to that considered by Lingen [1]. One of the designs considered by Lingen was a paired trial comparing two oral cancer screening modalities: 1) examination by a dentist using a visual and tactile oral examination, and referral for biopsy only for frank cancers (Test 1); and 2) examination by a dentist using a visual and tactile oral exam, a second look with the VELscope oral cancer screening device, and stringent instructions to biopsy any lesion detected during either examination (Test 2).

We could find no published oral cancer screening trials of paired continuous tests. Instead, we chose parameter values from a breast cancer screening study [3] and an oral cancer screening demonstration study [20]. We fixed the sample size at 50,000 [3] and the rate of visible lesions at 0.1 (rate of suspicious oral cancer and precancerous lesions reported in 28 studies between 1971 and 2002 ranges from 0.02 to 0.17, Table 6, [20]). We approximated the disease prevalence as 0.01 based on the number of Americans with cancer of the oral cavity and pharynx [21] and the 2011 population estimate from the U.S. Census Bureau [23]. For the purposes of the illustration, the *true* areas under the curves for Test 1 and Test 2 were fixed at 0.77 and 0.71, respectively, similar to the areas under the curves reported for digital and film mammography [3]. In truth, we have no data to support the superiority of the visual tactile oral exam over the visual tactile oral exam plus scope.

The trial by Pisano *et al.* compared digital and film mammography [3]. Since the recall rate for both modalities was 8.4%, the trial was unlikely to be biased. In the hypothetical oral cancer screening trial, however, we posit that there would be a large difference in the percent ascertainments for each screening modality. In the first arm, the dentist only recommends biopsy for participants with highly suspicious lesions. Thus, we fixed the percent ascertainment to be very low, only 0.01%. The oral pathologist recommends biopsy for almost any lesion so we set the percent ascertainment at 97%. The large difference in percent ascertainment between the two screening tests creates extreme paired screening trial bias.

Under conditions of extreme bias, our bias correction method performs well, as shown in Figure 5. The *true* difference in the areas under the curves is 0.06 (p = 0.001), the *observed* difference is $-$ 0.06 (p = 0.005), and the *corrected* difference is 0.06 (p = 0.001). In the *observed* analysis, the receiver operating characteristic curves have the opposite orientation to the true state of nature. The *corrected* analysis removes the bias and the curves are realigned with the true state of nature.

In reality, the study investigator would not know which analysis had results closest to the truth. To validate our choice of analysis, we simulated the hypothetical study design using the parameter values specified above. The simulated Type I error rate of the corrected analysis is below nominal at 0.03, while the Type I error rate of the observed analysis is above nominal at 0.06. The correct rejection fraction of the bias-corrected analysis is 0.58, while that of the observed analysis is zero. In fact, using the observed analysis, the study investigator would wrongly conclude that Test 2 is superior to Test 1 86% of the time. Based on this simulation, we would recommend the study investigator use the bias correction method to correct the results of the hypothetical oral cancer screening study.

**Discussion**

We could find no other methods that attempted to ameliorate paired screening trial bias for paired cancer screening studies that use the difference in the full area under the curve as a metric. Re-weighting, generalized estimating equations, imputation and Bayesian approaches have been proposed to reduce the effect of partial verification bias [*e.g.*, 11, 24 $-$ 29]. Maximum likelihood methods [30] and latent class models [31] have been proposed to estimate diagnostic accuracy in the presence of imperfect reference standard bias.

The amount of bias in the study and the performance of the bias correction method depend on fourteen factors: the means, variances, and correlation of the case and non-case test scores, the disease prevalence, the rate of signs and symptoms, and the percent ascertainment for each screening test. After careful analysis of over $170,000$ combinations of the parameter values, we were unable to determine a definitive pattern upon which to base recommendations for using the standard analysis versus a bias-corrected approach.

In order to determine if bias correction is indicated for a specific screening trial, we recommend the investigator conduct a simulation study similar to those described in the manuscript. The simulation software will be made available at www.samplesizeshop.com. The software will simulate Type I error rate and power for both the standard and bias-corrected analyses. The study investigator should choose the analysis that has the Type I error rate closest to the nominal level and the highest correct rejection fraction.

The software will request that the study investigator specify the values of each of the fourteen factors. Estimates can be based on previous studies or auxiliary information. For instance, the rate of signs and symptoms is usually not described in the literature. In the Results section, we calculated an estimate for the rate of signs and symptoms using surrogate values published in two breast cancer screening studies. When convincing values cannot be found, the study investigator can conduct a sensitivity analysis using a range of values to see how strongly that parameter affects the choice of analysis.

In this article, we treat the prevalence of disease as a fixed feature of our study population. However, the disease prevalence may be higher or lower for different subgroups of the population. For instance, in the DMIST study, women younger than 50 years of age had a much lower prevalence of invasive breast carcinoma than women greater than 50 years of age [3]. At this time, the method does not allow for a heterogeneous case mixture. To allow for differences in disease prevalence in the study population, the study investigator can stratify participants into subgroups with a homogeneous disease prevalence. The investigator would then conduct separate analyses for each subgroup.

In some instances, both the standard analysis and the corrected analysis will have higher than nominal Type I error rates. In this case, we recommend that investigators avoid using a paired screening study design. Another option would be to randomize study participants to one of two possible screening modalities. However, a randomized trial is inefficient relative to a paired screening trial, and may still be subject to differential verification bias. A more clinically useful endpoint may be to consider mortality, as was done in the National Lung Screening Trial (NLST) [32]. Unlike receiver operating characteristic analysis, mortality trials such as the NLST are not vulnerable to paired screening trial bias, or verification bias in general.

The bias correction method is a maximum likelihood method. Thus, the accuracy of the estimation depends on the number of cases. We do not recommend using the method for studies with a very small number of cases ($< 500$), and interval cases ($< 5$).

This paper provides two contributions to the statistical literature. First, we describe a method to correct for paired screening trial bias, a bias which has not been addressed by other correction techniques. Due to the increasing use of continuous biomarkers for cancer (see, *e.g.*, [33]), a growing number of screening trials have the potential to be subject to paired screening trial bias. The proposed method will counteract bias in the paired trials and allow investigators to compare screening tests with fewer decision errors. Second, we introduce an important metric for evaluating the performance of bias correction techniques, that of reducing decision errors. We recommend that every new bias correction method be evaluated with a study of Type I error and power.

## Conclusions

The proposed bias correction method reduces decision errors in the paired comparison of the full area under the curve of screening tests with Gaussian outcomes. Because the performance of the bias correction method is affected by characteristics of the screening tests and the disease being examined, we recommend conducting a simulation study using our free software before choosing a bias-corrected or standard analysis.

## Competing Interests

The authors declare that they have no competing interests.

**Authors' contributions**

BMR conducted the literature review, derived the mathematical results, designed and programmed the simulation studies, interpreted the results, and prepared the manuscript. TAA assisted with the literature review and provided expertise on the context of the topic in relation to other work in the field. JTB assisted with the mathematical derivations. SMK provided guidance for the design and programming of the simulation studies. AM improved the software and packaged it for public release. KEM reviewed the intellectual content of the work and gave important editorial suggestions. DHG conceived of the topic and guided the development of the work. All authors read and approved the manuscript.

## Acknowledgements

The research presented in this paper was supported by two grants. The mathematical derivations, programming of the algorithm and early simulation studies were funded by the NCI 1R03CA136048-01A1, a grant awarded to the Colorado School of Public Health, Deborah Glueck, Principal Investigator. Completion of the simulation studies, including the Type I error and power analyses, was funded by NIDCR 3R01DE020832-01A1S1, a minority supplement awarded to the University of Florida, Keith Muller, Principal Investigator, with a subaward to the Colorado School of Public Health. The content of this paper is solely the responsibility of the authors, and does not necessarily represent the official views of the National Cancer Institute, the National Institute of Dental and Craniofacial Research, nor the National Institutes of Health.

**Table 1.** Quadrant definitions.

| Quadrant Name | Definition |
|:---:|:---:|
| $Q_1$ | $\{x_{i1k} \geq a_1; x_{i2k} \geq a_2\}$ |
| $Q_2$ | $\{x_{i1k} \geq a_1; x_{i2k} < a_2\}$ |
| $Q_3$ | $\{x_{i1k} < a_1; x_{i2k} \geq a_2\}$ |
| $Q_4$ | $\{x_{i1k} < a_1; x_{i2k} < a_2\}$ |

**Table 2.** Effect of percent ascertainment on the Type I error rate. Type I error rates are calculated over 10,000 repetitions of the data for the hypothesis test of a difference in the full areas under the curves. The nominal Type I error is fixed at 0.05.

| Paired Screening Trial Bias? | Disease Prevalence | Percent Ascertainment (Test 1 / Test 2) | True | Observed | Corrected |
|---|---|---|---|---|---|
| Yes | 0.01 | 15/50 | 0.01 | 0.89 | 0.36 |
|  | 0.01 | 15/80 | 0.02 | 0.95 | 0.25 |
|  | 0.01 | 50/80 | 0.01 | 0.23 | 0.12 |
|  | 0.14 | 15/50 | 0.02 | 1.00 | 0.82 |
|  | 0.14 | 15/80 | 0.02 | 1.00 | 0.60 |
|  | 0.14 | 50/80 | 0.02 | 1.00 | 0.20 |
|  | 0.24 | 15/50 | 0.02 | 1.00 | 0.95 |
|  | 0.24 | 15/80 | 0.02 | 1.00 | 0.91 |
|  | 0.24 | 50/80 | 0.02 | 1.00 | 0.40 |
| No | 0.01 | 15/15 | 0.01 | 0.02 | 0.23 |
|  | 0.01 | 50/50 | 0.01 | 0.02 | 0.12 |
|  | 0.01 | 80/80 | 0.02 | 0.02 | 0.18 |
|  | 0.14 | 15/15 | 0.02 | 0.02 | 0.26 |
|  | 0.14 | 50/50 | 0.02 | 0.02 | 0.14 |
|  | 0.14 | 80/80 | 0.02 | 0.02 | 0.03 |
|  | 0.24 | 15/15 | 0.02 | 0.02 | 0.26 |
|  | 0.24 | 50/50 | 0.02 | 0.02 | 0.14 |
|  | 0.24 | 80/80 | 0.02 | 0.02 | 0.04 |

**Figure 1.** Flowchart of a paired trial of two continuous screening tests. The flowchart culminates in the study investigator's observation of the disease status of the participant.

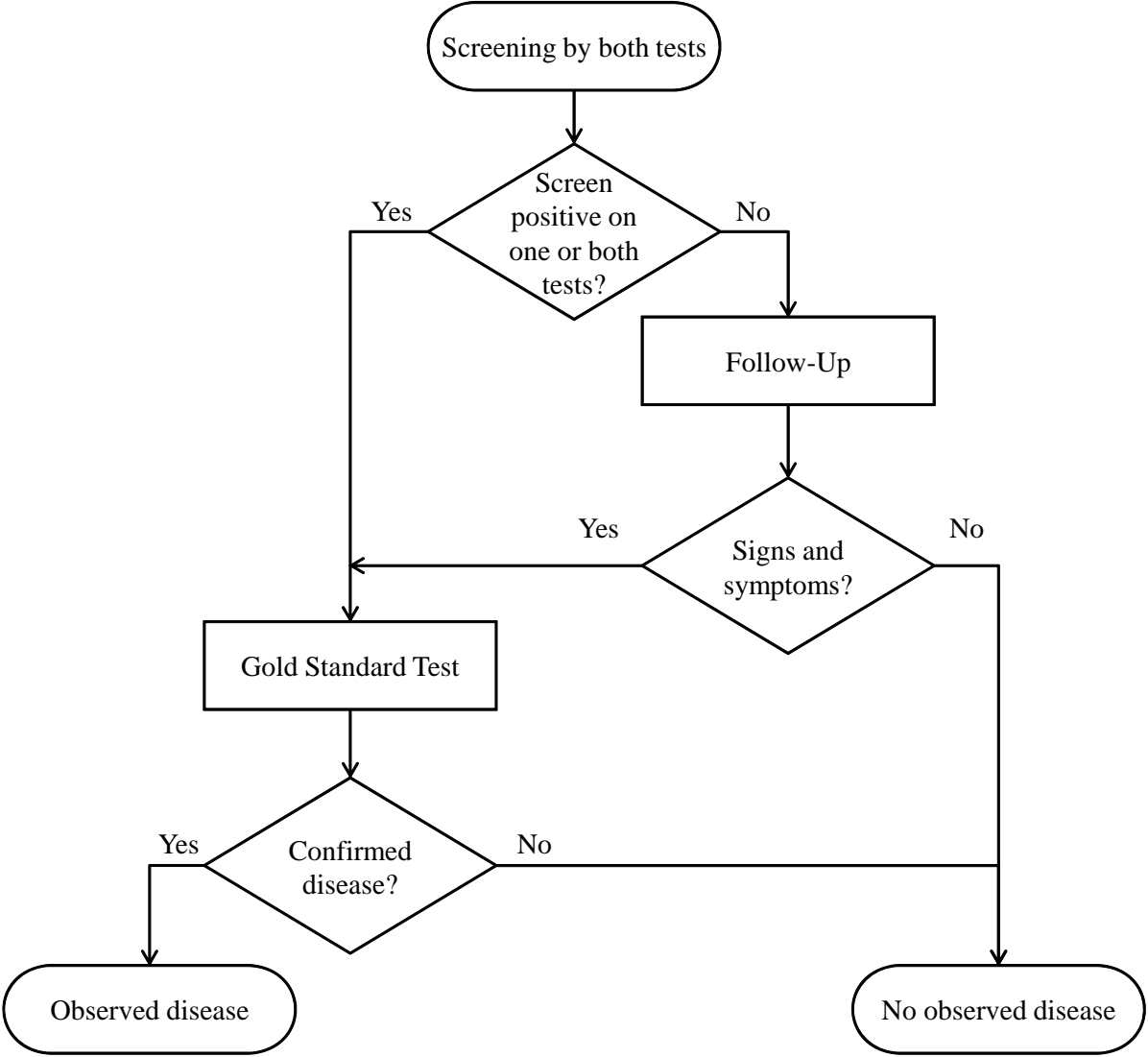

**Figure 2.** Hypothetical data for a paired screening trial. Paired screening test scores are for the non-cases, screen-detected cases, interval cases, and missed cases in a hypothetical screening study comparing two continuous screening tests.

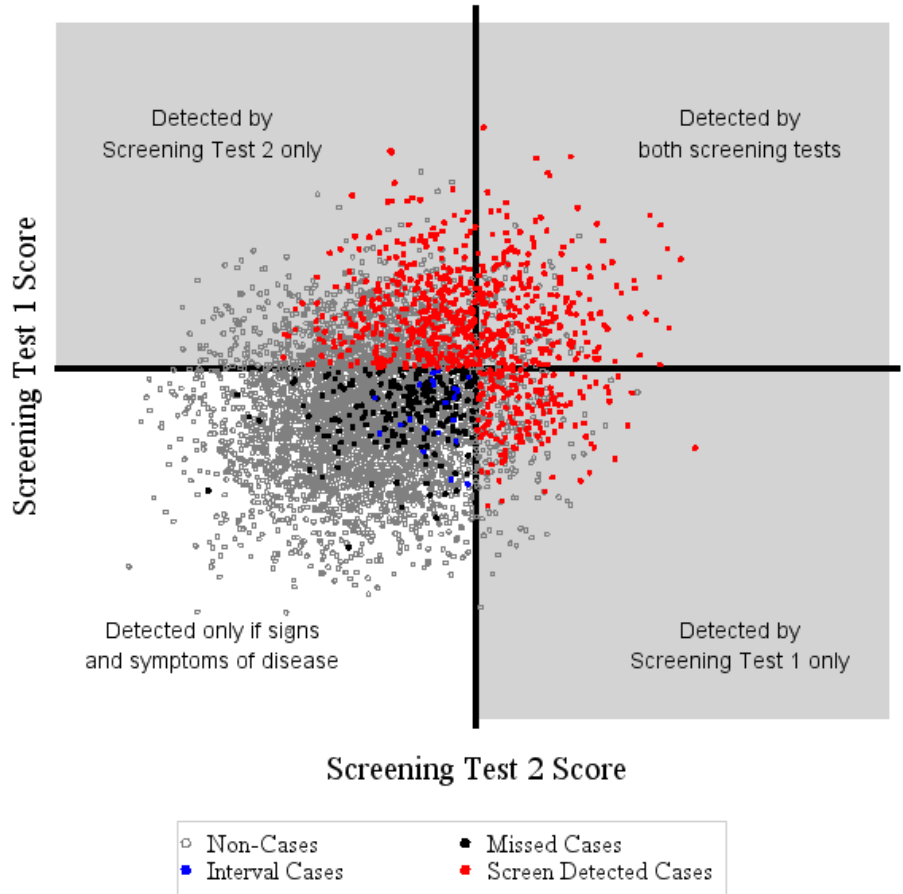

**Figure 3.** Effect of disease prevalence on Type I error rate and power. The nominal Type I error was fixed at 0.05. Graphs show the proportion of times the hypothesis test rejects and A) the null holds, B) the alternative holds and the conclusion of the hypothesis test agrees with the true state of nature, or C) the alternative holds but the conclusion of the hypothesis test is opposite the true state of nature.

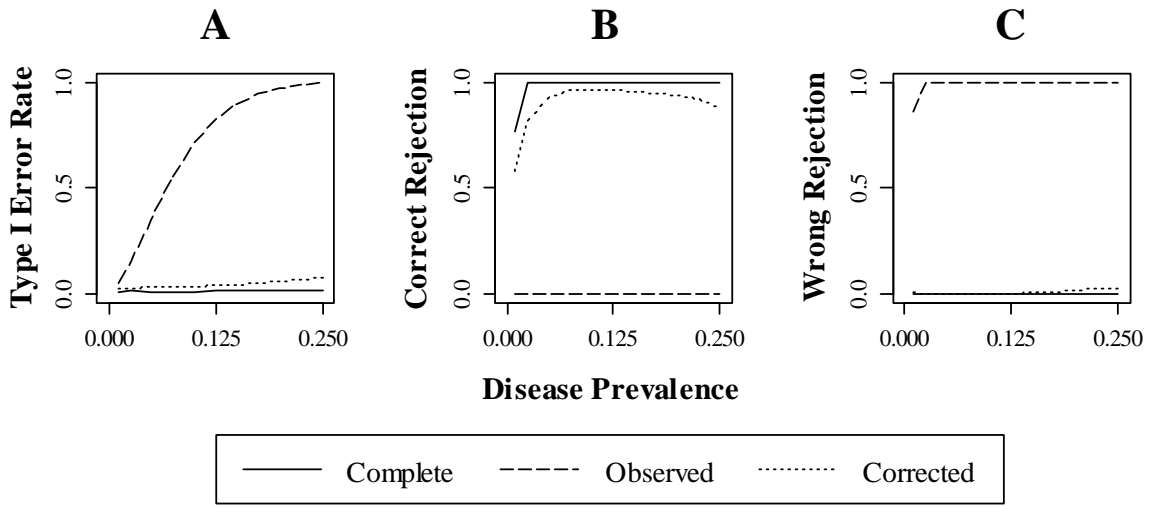

**Figure 4.** Effect of the rate of signs and symptoms on Type I error rate and power. The nominal Type I error was fixed at 0.05. Panels A, B, and C are as in Figure 3.

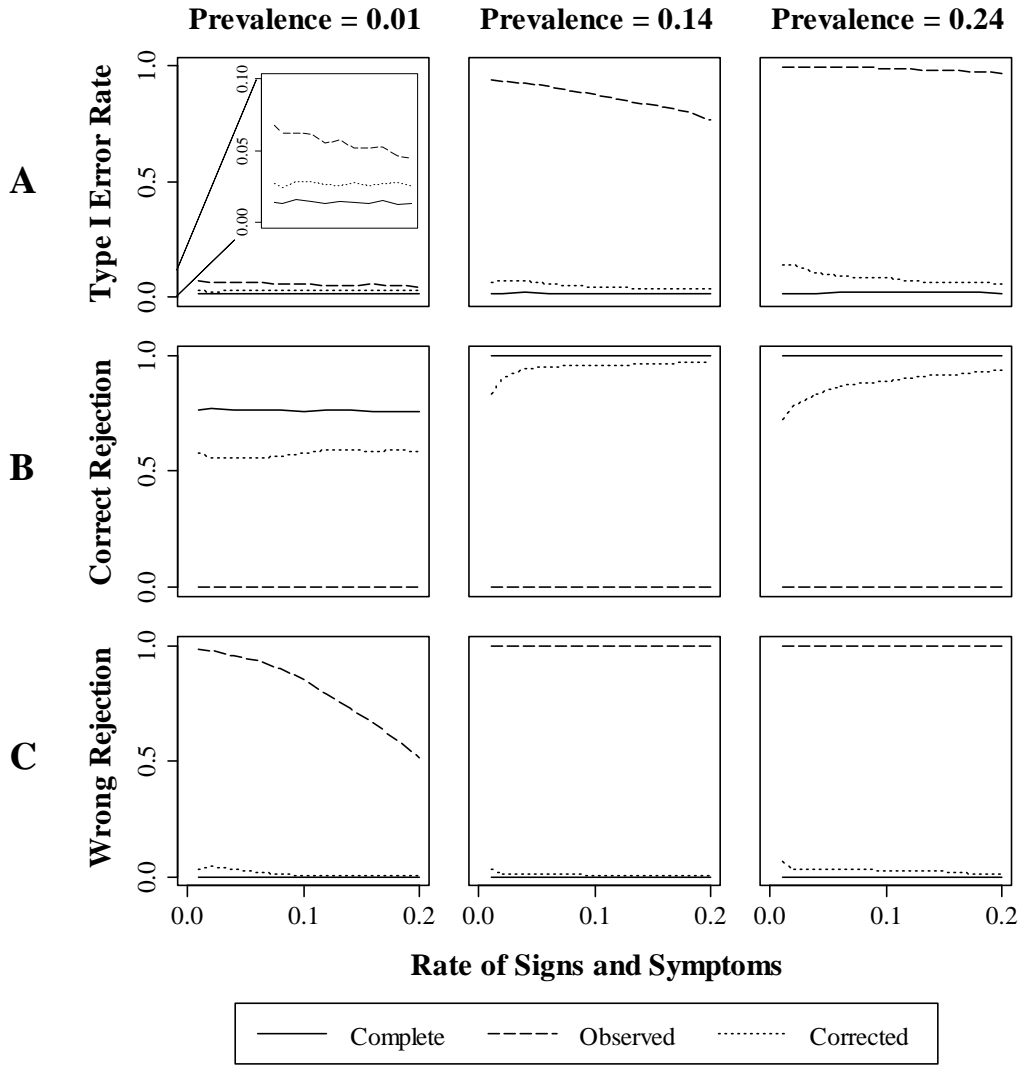

**Figure 5.** Receiver operating characteristic curves for a hypothetical oral cancer screening study. The study is subject to paired screening trial bias. The *true* areas under the curves for Test 1 and Test 2 are 0.77 and 0.71, respectively, for a *true* difference of 0.06. The *observed* difference is $-0.06$, with the *corrected* difference at 0.06.

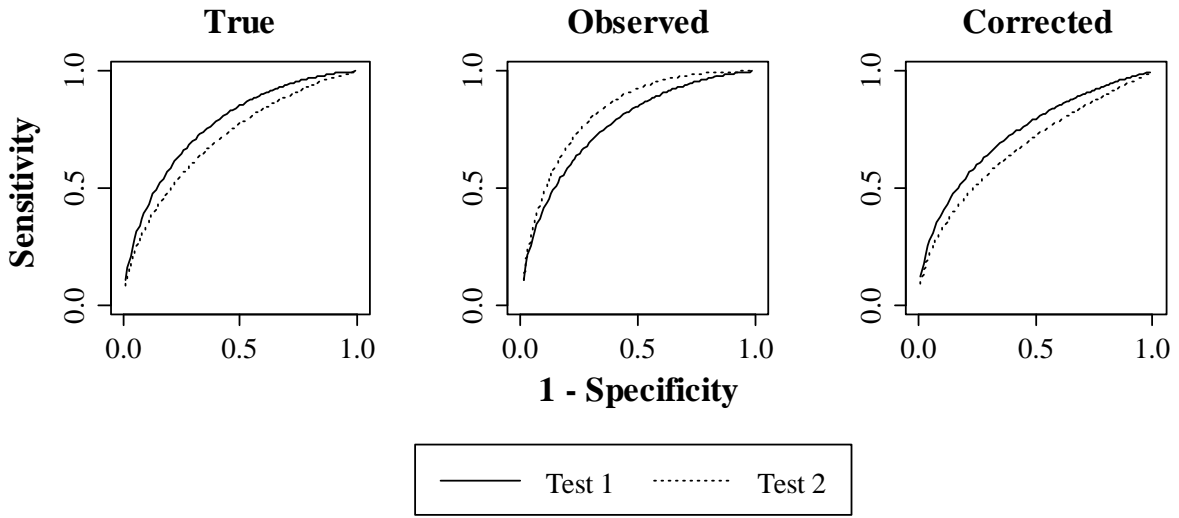